\documentstyle [12pt,dina4] {article} 
\title{From stochastic differential equations
          to quantum field theory}

\author{\underline{R. Gielerak}\thanks{also: Technical University of
Zielona G{\.o}ra, Inst. of T. Phys., Poland}
 \\ gielerak@ift.uni.wroc.pl \\ [2ex]
P. \L ugiewicz  \\piotrlug@ift.uni.wroc.pl \\  [2ex]
Institute of Theoretical Physics\\University of Wroc\l aw\\
50-204 Wroc\l aw, Poland}
\begin{document}
\maketitle
\begin{abstract}
 Covariant stochastic partial (pseudo)-differential equations are 
studied 
in
any dimension. In particular a large class of covariant interacting 
local
quantum fields obeying the Morchio-Strocchi system of axioms for 
indefinite
quantum field theory is constructed by solving the analysed equations.
The associated random cosurface models are discussed and some 
elementary
properties of them are outlined.
\end{abstract}


\section{Introduction}
The strict connection of the Euclidean (bosonic) quantum field theory 
with 
(infinite-dimensional) Stochastic Analysis objects and concepts 
\cite{dPZ1, dPZ2} is well known in different situations. Let us 
recall few 
of them. \\

{\bf Example 1} {\bf Scalar fields as solutions of the Ito -type 
SDEs}\\
Let $B_t$ be a cylindric version of ${\cal S}'({\bf R}^d)$-valued 
Brownian motion (where
${\cal S}'({\bf R}^d)$ stands for the space of real tempered 
distributions)
i.e. for any $f \in {\cal S}({\bf R}^d)$ the coordinate process $b^f_t
\equiv (B_t,f)$ is a version of one-dimensional Brownian motion. The 
linear
Ito equation
\begin{equation}
\label{eq1}
d\xi _t^0 = \sqrt{1- \Delta} \xi ^0_t dt  + dB_t
\end{equation}
has a \underline{stationary} solution $\xi_t^0$ which has a law 
${\cal L}
(\xi_t^0)=\mu _0^N$ easily identified with the free Nelson field, i.e.
$\mu_0^N$ is a Gaussian probability measure on
$({\cal S}'({\bf R}^{d+1}), {\cal B}({\cal S}'({\bf R}^{d+1}))$
$\equiv$ cylinder $\sigma$
-algebra) ($\equiv$ Gaussian generalized
random field) with the mean equal to zero and covariance
\begin{equation}
\label{eq2}
{\bf E} \xi^0_0({\bf x}) \xi^0_t({\bf y}) = \frac{\hbox{e}^{-
|t|\sqrt{1-
\Delta}}}{2\sqrt{1- \Delta}}({\bf x}-{\bf y})= \int_{{\cal S}'({\bf 
R}^{d+1})}d\mu_0^N (\varphi) \varphi(0,{\bf x}) \varphi (t,{\bf y}).
\end{equation}
In particular, the invariant measure $\nu _0 ^{(N)}$ of the
${\cal S}'({\bf R}^{d})$-valued Markov diffusion $\xi_t^0(x_0))$
(where $x_0 \in {\cal S}'({\bf R}^{d}$ is the initial condition)
corresponding to \ref{eq1} is easily computable with the result that
$\nu_0^{(N)}$ is centered Gaussian measure on
$({\cal S}'({\bf R}^{d}), {\cal B}({\cal S}'({\bf R}^{d})))$ 
with the following characteristic
functional:
\begin{equation}
\label{eq3}
\int_{{\cal S}'({\bf R}^{d})} d\nu_0^{(N)}(\varphi) 
\hbox{e}^{i(\varphi,f)}=
\hbox{e}^{-\frac{1}{2}||f||_{-\frac{1}{2}}^2}  
;
\end{equation}
where
$$
  ||f||^2_{-\frac{1}{2}} = \int  f({\bf 
x})(1-\Delta)^{-\frac{1}{2}}({\bf x}-{\bf y})f({\bf y}) d{\bf x}d{\bf 
y}.$$
For a various aspects of the process(es) $\xi_t^0$ (resp. 
$\xi_t^0(x_0)$)
we refer to \cite{AHK1, Ro1, AGR1, AGR2}.

{\bf Example 2} {\bf Scalar fields as invariant measures of the
Ito -type SDEs.}\\
Let us consider the following stochastic differential equations:
\begin{equation}
\label{eq4}
dL_t^{\epsilon ,0}= (1-\Delta)^{\epsilon}L_t^{\epsilon ,0} dt +
                    (1-\Delta)^{\frac{\epsilon - 1}{2}} dB_t
\end{equation}
where $dB_t$ is the cylindric ${\cal S}'({\bf R}^{d+1})$-valued 
Brownian
motion and the (regularizing) parameter $\epsilon \in (0,1]$. By 
simple
computations it follows that the invariant measure for all of the 
equations
(\ref{eq4}) is equal to the free Nelson field $\mu_0^{(N)}$. The law
${\cal L}(L_t^{1,0}) \equiv \mu_0^{(L)}$ of the \underline{stationary}
solution of (\ref{eq4}), called the free Langevin field, is easily 
seen as
centered Gaussian probability measure on
$({\cal S}'({\bf R}^{d+1}), {\cal B}({\cal S}'({\bf R}^{d+1})))$
characterized by
\begin{equation}
\label{eq5}
\int_{{\cal S}'({\bf R}^{d+1})} d\mu_0^{(L)}(\varphi) 
\hbox{e}^{i(\varphi,f)}=
\hbox{e}^{-\frac{1}{2}||f||_{-1,-2}^2}
;
\end{equation}
where
$$
  ||f||^2_{-1,-2} = \int  f(s,x)(-\partial_0^2+(1-\Delta)^2)^{-1}(s-
t,x-y)
  f(t,y)dsdxdtdy.
$$
The corresponding to $d\mu_0^{(L)}$ generalized random field is
(sharp)-Markov in the computer time direction $t$  and (germ)-Markov 
in 
the other directions.

The study of the equations \ref{eq4} is a part of the so called 
stochastic
quantisation programme \cite{PaWu, JLM, Hu, dPT, RZ1, RZ2}.

{\bf Example 3} {\bf Free quantum fields as stochastic integrals.}\\
{\bf 3.A}   Let $\eta$ be a Gaussian white noise on the space
${\cal S}'({\bf R}^{d})$, i.e. $\eta$ is distributed according to the
Gaussian measure $d\mu^{GN}$ characterized by:
\begin{equation}
\label{eq6}
\int_{{\cal S}'({\bf R}^{d})} d\mu^{GN}(\eta) \hbox{e}^{i(\eta,f)}=
\hbox{e}^{-\frac{1}{2}||f||_{2}^2}.
\end{equation}
For $\lambda \in (0,\frac{1}{2}]$, let us consider the following 
partial
(pseudo)-differential stochastic equation
\begin{equation}
\label{eq7}
(1-\Delta)^{\lambda} \varphi _{\lambda} = \eta.
\end{equation}
The solution of (\ref{eq7}), given by the stochastic integral
\begin{equation}
\label{eq8}
\varphi _{\lambda} = (1- \Delta)^{-\lambda} \ast \eta
\end{equation}
is easily recognized as a generalized free (Euclidean) quantum field 
with
the two-point Schwinger function
\begin{equation}
\label{eq9}
{\bf E} \varphi_{\lambda}(x)\varphi_{\lambda}(y) = (1-\Delta)^{-2 
\lambda}
(x-y)
\end{equation}
In particular, for $\lambda = \frac{1}{2}$, $\varphi_{\lambda}$ is 
identical to the free Nelson field.

{\bf 3.B}    Now, let
$\eta$ be a ${\cal S}'({\bf R}^{d}) \otimes {\bf R}^N$-valued
Gaussian white noise and let $\tau$ be some real (orthogonal)
representation of rotation group $(S)O(d)$ in the space ${\bf R}^N$ 
and let
${\cal D}$ be $\tau$-covariant differential operator. Providing that 
${\cal
D}$ is such that corresponding Green function ${\cal D}^{-1}$ can be
properly defined, it follows that the stochastic integral $A \equiv 
{\cal
D}^{-1} \ast \eta$ gives the solution of the following covariant 
stochastic
differential equation
\begin{equation}
\label{eq10}
      \tilde{ {\cal D}} A = \eta.
\end{equation}
where $\tilde{{\cal D}}$ is the adjoint of ${\cal D}$ in the 
canonical 
pairing ${}_{{\cal S}'}<\cdot ,\cdot >_{{\cal S}}$. In particular, 
taking 
$d=3$, $\tau = {\bf D}_1 \oplus {\bf D}_1$ and
\begin{equation}
\label{eq11}
{\cal D} =      \left(
                \begin{array}{cccccc}
                m & 0 & 0 & 0 & b \partial_z & -b \partial _y \\
                0 & m & 0 & -b \partial_z & 0 & b \partial_x \\
                0 & 0 & m & b \partial_y & -b \partial_x & 0 \\
                0 & c\partial_z & -c\partial_y & m & 0 & 0 \\
                -c\partial_z & 0 & c\partial_x & 0 & m & 0 \\
                c\partial_y & -c\partial_x & 0 & 0 & 0 & m
                \end{array}
                \right)
\end{equation}
with $b^2 = c^2 = 1$, and $bc=-1$ it follows that the stochastic 
integral
${\cal D}^{-1}\ast \eta$ for $\eta$ being pure Gaussian white noise 
gives 
two independent copies of two real massive(with the mass $m$) 
Euclidean 
Proca fields.
For a systematic approach to such constructions see \cite{BGL} and 
for a
particular application to the free $EM_4$ fields \cite{AIK, AGW1, 
AGW2}.

The particular features of the above listed examples are: the 
linearity of 
the corresponding equations and Gaussianity of the corresponding 
noise.
These features lead to the Gaussian solutions, therefore not very 
interesting
from the point of view of physics. The interesting physics seems
to be described by non-Gaussian examples. In order to get them two 
different
approaches were introduced. The first approach is to perturb the 
(linear)
drifts by adding some nonlinear perturbation. However the main 
difficulty
here is that the typical realisations (sample paths) of the underlying
solutions are generically distributions (not functions!). The second 
approach
to the problem of constructing non-Gaussian examples is to change the
Gaussian noise into some tractable non-Gaussian noise. The simplest
possibility is to (perturb\slash exchange) the Gaussian white-noise
by Poisson noise.\\

{\bf Example 1} (continuation).\\
It is well known \cite{AHK1, ARZ, RZ1, RZ2} that at least for $d=1$ 
there
exist nonlinear
measurable perturbations of the linear drift in (\ref{eq1}) which 
lead 
to the
stationary solutions of the corresponding Ito SDE being identical (on 
the 
level of laws) to the interacting models of scalar fields constructed 
in the
so called Constructive Quantum Field  Theory (\cite{Sim, GJ} and 
references
therein).

However the problem is that the explicit form of the corresponding
perturbations is not known. Nevertheless this shows that there exist
nonlinear Ito SDEs on ${\cal S}'({\bf R}^d)$ that lead to nontrivial 
quantum
fields obeying all Wightman axioms. The challenging problem to 
describe 
such
possibilities in an explicit (or constructive) form is still open.

The attempts to perturb the Gaussian noise $dB_t$ in \ref{eq1} and 
\ref{eq7} 
by some non-Gaussian one are easily tractable, however the problem of
stability of the crucial reflection positivity of the solutions seems 
to have
negative solution \cite{AGW2} in general.\\
{\bf Example 2} (continuation).

At least for $d=2$ the nonlinear perturbations of \ref{eq4} of 
gradient 
type 
(but see also \cite{JLS} for nongradient type case)
were studied intensively \cite{PaWu, JLM, Hu, dPT, RZ1, RZ2}.
The typical, two dimensional
quantum field theory models like $P(\Phi)_2$ are again obtained as
stationary distributions of the underlying Markov diffusions. A new 
approach
to these equations, based on the idea of the "ground state 
transformation"
together with the methods of the Constructive Quantum Field Theory
\cite{GJ, Sim}
was recently invented in \cite{AGR}.

After preparing this report we get a copy of \cite{scat}in which some 
of the
results presented here are also obtained.

In the present exposition we shall focuss our attention on a recent 
progress
connected to the perturbation of the noise in the part {\bf 3B} of 
Examples 3.


\section{Interacting local covariant quantum fields from Covariant 
SPDEs}

Let $d\mu^{(P,\tau)}$ be a regular Poisson $\tau$-covariant noise on 
the
space ${\cal S}'({\bf R}^d)\otimes {\bf R}^{dim \tau}$, where $\tau$
is some real representation of the rotation group $SO(d)$, $d \geq 
2$. The
noise $d\mu^{(P,\tau)}$ is characterized by
\begin{equation}
\label{eq2.1}
\int_{{\cal S}'({\bf R}^d)\otimes {\bf R}^{dim \tau}}
\hbox{e}^{i(\varphi , f)} d \mu^{(P, \tau)}(\varphi)= \hbox{e}^{\int 
dx
 \int d \lambda (\alpha) (\hbox{e}^{i<\alpha , f(x)>}-1)}
\end{equation}
where we assume that the so-called Levy measure $d\lambda$ on ${\bf
R}^{dim \tau}$ is such that \\
(i) $d\lambda$ has all moments \\
(ii)$d \lambda $ is $\tau$-invariant

Let $\tau '$ be an another real representation of the group $SO(d)$ 
in 
${\bf R}^{dim \tau '}$. Here, for simplicity we assume that
dim$\tau$=dim$\tau '$ reffering to a more general case
dim$\tau \neq$ dim$\tau '$ to our (forthcoming) paper \cite{GL}. The 
action
$\tau$ (resp. $\tau '$) of the group $SO(d)$ can be naturally lifted 
to the
action $T_{\tau}$ (resp. $T_{\tau '}$) of $SO(d)$ in the space
${\cal S}'({\bf R}^d)\otimes {\bf R}^{dim \tau}$.

Recall that a first order (for simplicity again) differential operator
${\cal D} = \sum_{i=1}^{d}B_i \partial _i + M$, where $B_i$, $M \in 
Hom({\bf R}^{dim \tau}; {\bf R}^{dim \tau '})$ is called $(\tau,\tau
')$-covariant iff the following diagram
\begin{equation}
\label{eq2.2}
\begin{array}{ccc}
{\cal S}'({\bf R}^d)\otimes {\bf R}^{dim \tau}
& \quad{\buildrel {\cal D} \over\longrightarrow}\quad &
{\cal S}'({\bf R}^d)\otimes {\bf R}^{dim \tau '} \\
T_{\tau} \Big\downarrow & & \Big\downarrow T_{\tau '} \\
{\cal S}'({\bf R}^d)\otimes {\bf R}^{dim \tau}
& \quad{\buildrel {\cal D} \over\longrightarrow}\quad &
{\cal S}'({\bf R}^d)\otimes {\bf R}^{dim \tau '}
\end{array}
\end{equation}
do commute.

Let us denote by $Cov(\tau, \tau ')$ the set of all such $(\tau, \tau
')$-covariant differential operators. The complete description of the 
sets
$Cov(\tau, \tau ')$ is given in the paper \cite{GL}. From the 
definition 
of ${\cal D} \in Cov(\tau, \tau ')$ it follows that the symbol
$\sigma_{{\cal D}}$ of ${\cal D}$ defined as $\sigma _{{\cal D}}(p) 
\equiv
i \sum_{j=1}^{d} B_j p_j$ has the property:
$$
\hbox{det}(\sigma _{{\cal D}}(p) + m {\bf 1})= c \prod _{k=1}^{n} 
(p_1^2+... p_d^2 + m_k^2)
$$
where $m_k \in {\bf C}$, $k=1,...,n$ , $n \leq N \slash 2$, c is a 
complex
number.
If all $m_k$
are real and $c \neq 0$ the operator ${\cal D}$ is invertible on 
suitably
chosen function space and in this case we shall call it
\underline{admissible}. If additionally
all $m_k \neq 0$, operator ${\cal D}$ is said to
have a strictly positive mass spectrum.

We shall consider SPDEs of the type:
\begin{equation}
\label{eq2.3}
\tilde{{\cal D}} \varphi = \eta
\end{equation}
where: $\eta$ is given (regular) $\tau$-covariant noise on
${\cal S}'({\bf R}^d)\otimes {\bf R}^{dim \tau}$, ${\cal D} \in
Cov(\tau , \tau ')$ is such that the Green function ${\cal D}^{-1}$ 
of 
${\cal D}$ can be defined as a continous imbedding of some nuclear 
space
${\cal F}\otimes{\bf R}^{dim \tau '}$ into the space
${\cal S}'({\bf R}^d)\otimes {\bf R}^{dim \tau}$ (such operators are
called regular).

A generalized random field $\varphi$ indexed by
${\cal F}\otimes{\bf R}^{dim \tau '}$ is called weak solution of
(\ref{eq2.3}) iff $(\varphi, f) \cong (\eta , {\cal D}^{-1}f)$ for 
all 
$f \in {\cal F}\otimes{\bf R}^{dim \tau '}$ where $\cong$ means the 
equality 
inlaw. Let $\Gamma _{\eta}$ denote the characteristic functional of 
the 
field
$\eta$, then the characteristc functional $\Gamma _{\varphi}$ of the 
weak
solution of (\ref{eq2.3}) is given by: 
$\Gamma_{\varphi}= \Gamma_{\eta}({\cal D}^{-1}f)$. 
In particular, if $\eta$ 
is a $\tau$-covariant regular whitenoise then the characteristic 
functional 
od $\varphi$ is given by:\begin{equation}
\label{eq2.4}
\Gamma_{\varphi}(f)= 
\hbox{e}^{-\frac{1}{2}<{\cal D}^{-1}f|A{\cal D}^{-1}f>}
\hbox{e}^{\int dx 
\int d \lambda(\alpha) (\hbox{e}^{i<\alpha , {\cal D}^{-1}
f(x)>}-1)}.
\end{equation}
From the assumption that $d\lambda$ has all moments it follows that 
all
Schwinger functions of the field $\varphi$ do exist.

The main observation is the following:\\
{\bf Theorem 1}({\bf Existence of Wightman functions})\\
{\it Let
$\eta$ be a regular, $\tau$-covariant white noise and let ${\cal D} 
\in
Cov(\tau , \tau ')$ has an admissible mass spectrum. By
$\tilde{\tau '}$ we denote the analiticaly continued real 
representation
$\tau '$ to the corresponding (real) representation of the
special orthochronous Lorentz group. Then, there exists
a system of
tempered distributions ${\cal W}_n^{\tilde{\tau '}}$ which is: local,
covariant
(with respect to $\tilde{\tau '}$),
spectral and such that restrictions of the moments of the field 
$\varphi$
being a weak solution of ${\cal D} \varphi = \eta$ to the set
$x_1^0 <...<x_{n+1}^0$ are equal to the Laplace-Fourier 
transformations of
certain linear combinations ${\cal W}_n^{{\tau '}}$ of
${\cal W}_n^{\tilde{\tau '}}$, i.e.: for $x_1^0 <...<x_{n+1}^0$}
$$
{\bf E} \varphi (x_1^0,{\bf x}_1)...\varphi (x_{n+1}^0,{\bf x}_{n+1})=
$$
$$
=\int \hbox{e}^{-\sum_{j=1}^n p_j^0(x_{j+1}^0-x_j^0)}
\hbox{e}^{i\sum_{j=1}^n {\bf p}_j({\bf x}_{j+1}-{\bf x}_j)}
{\cal W}_n^{\tau '}(p_1,...,p_n) \otimes _{j=1}^n dp_j
$$

{\bf Proof:} See \cite{BGL}, \cite{scat}.\rule{5pt}{5pt}

The very interesting question whether there exist non-Gaussian 
examples of
such equations which lead to reflection positive solutions is still 
unsolved (although there
are strong negative indications) . However, in the contex of
nonpositive quantum field theory axiomatized by Morchio-Strocchi in
\cite{MoSt}
the constructed models may lead to interesting new examples of
interacting quantum fields with nontrivial scatering matrices.

{\bf Theorem 2} ({\bf Hilbert Space Structure Condition})\\
{\it 
Assume that $\eta$ as in Theorem 1,
$\;\;{\cal D} \in Cov(\tau , \tau ') \;\;$
with the admissible mass spectrum \\
$\{ m_1,...,m_n \}$ such that
$m_l \neq m_j$ for $l \neq j$. Then there exists a sequence 
$\{ || \cdot ||_n
\}$ of Hilbert norms on
${\cal S}'({\bf R}^d)\otimes {\bf R}^{dim \tilde{\tau '}}$ which are
continous in the Schwartz topology and such that}
$$
|{\cal W}_{m+n}(f_m^{\ast}\otimes g_n)| \leq ||f||_m ||g||_n
$$
{\it   
for all $f \in {\cal S}'({\bf R}^{dm})\otimes 
{\bf R}^{dim \tilde{\tau '}}$,
$g \in {\cal S}'({\bf R}^{dm})\otimes {\bf R}^{dim \tilde{\tau ' }}$
and where $f^{\ast}$ is an appropriate conjugation of $f$.}

{\bf Proof:} It follows by elaborating on the explicit form of the
corresponding Wightman functions $\{ {\cal W}_n^{\tilde{\tau '}} \}$
as obtained in
\cite{BGL}. For details we refer to \cite{GL}.\rule{5pt}{5pt}

The important consequence of this Theorem is that the corresponding 
GNS
inner-product space obtained from $\{ {\cal W}_n^{\tilde{\tau '}} \}$
has a natural structure
of a Krein space. This means that the underlying infrared
singularities are not
so bad; see \cite{MoSt};see also \cite{AGW1, AGW2} for a related 
models.

{\bf Remark}\\
{\it The assumption $m_l \neq m_j$ for $l \neq j$ in the Theorem 2 is
not necessary. In general, in the presence of the Poisson part in the 
noise
one can impose an algebraic condition (see \cite{GL})
(on the covariant operator
${\cal D}$) which is sufficient to get the HSSC.
\section{Random cosurfaces}

Let $C_k$ denotes the set of ${\cal C}^1$-piecewise cocycles in 
${\cal R}^4$
, i.e. elements of $C_k$ are $k$-dimensional ($k=1,2,3$) ${\cal
C}^1$-piecewise boundaryless compact submanifolds of ${\cal R}^4$.
Let ${\cal D} \in Cov(\tau , \tau ')$,
where we assume that $\tau '$ contains for a fixed
$k=1,2,3$ at least one subrepresentation $\tau_{(k)} \subset \tau '$ 
which is:
for $k=1$ of vector type ($\tau _{(1)} \simeq (0,2)$),
for $k=2$ of skew-symetric tensor type, i.e.
$\tau _{(2)} \simeq (-1,2)\oplus (1,2) $
and $\tau _{(3)}$ is of skew-symetric tensor
type i.e. $\tau _{(3)} \simeq  (0,2)$
For regular ${\cal D}$ as above we consider the equation
\ref{eq2.3} and let $A^{(k)}$ be a part of the multiplet $\varphi$
transforming
covariantly under the subrepresentation $\tau_{(k)}$. Then, for a 
fixed
collection $\Gamma_1,...,\Gamma_n \in C_k$ we would like to give a
mathematical meaning to the following random map:
\begin{equation}
\label{eq3.1}
{\cal S}'({\bf R}^4)\otimes {\bf R}^{dim \tau _{(k)}} \ni A^{(k)}
\longrightarrow
\hbox{e}^{i A^{(k)}(\Gamma_1)}...\hbox{e}^{i A^{(k)}(\Gamma_n)}
\end{equation}
where
\begin{equation}
\label{eq3.2}
A^{(k)}(\Gamma _j) = \oint _{\Gamma_j} A^{(k)}
\end{equation}
where the right-hand side of (\ref{eq3.2}) is understood in the sense 
of
differential forms calculus. The map \ref{eq3.1} is called random 
cosurface
connected to the field $A^{(k)}$.\\

{\bf Proposition 1}\\
{\it
Let $(\tau' , \tau)$ be such that $\tau ' \supset \tau_{(k)}$ for some
some $k \in
\{1,2,3 \}$, $\eta$ is $\tau$-covariant regular pure Poisson noise; 
${\cal
D}\in Cov(\tau , \tau ')$ is regular and such that}
\begin{equation}
\label{spec}
{\cal D}^{-1}(x) \sim  \frac{1}{|x|^{4+\delta}}
~~~\hbox{for}~~|x| \rightarrow + \infty
\end{equation}
{\it
with $\delta >0$. Fix a collection $\{ \Gamma_1,...,\Gamma_{n} \}
\subset C_k$.
Then for almost every realisation of the field $A^{(k)}\in {\cal 
S}'({\bf
R}^{d})\otimes {\bf R}^{dim \tau_{(k)}}$ the random cosurface map
$\prod_{p=1}^{n}\hbox{e}^{i A^{(k)}(\Gamma_p)}$ is well defined and 
moreover
the following a.s. version of the Stokes Theorem is valid:}
\begin{equation}
\label{eq3.3}
A^{(k)}(\Gamma) = d A^{(k)} (\delta \Gamma)
\end{equation}
{\it
where $\delta \Gamma$ is the coboundary of $\Gamma$ and the equality
\ref{eq3.3} holds for almost every realisation of $A^{(k)}$}

The proof of this Proposition follows straithforwardly from
the following two technical
lemmas (the proofs of which are contained in \cite{GL}). To formulate
them let us recall that the set
$$
\{ A^{(k)}= \sum_{j} \alpha_j {\cal D}^{-1}|_{(k)}(x-x_j)~~|~~ \{ x_j 
\}
\hbox{is locally finite subset of}~{\bf R}^{d}; ~ \alpha \in 
\hbox{supp}d
\lambda \}
$$
is of measure $1$.\\

{\bf Lemma 1}\\
{\it Let} $\Sigma \subset {\bf R}^4$
{\it be of Lebesgue measure zero and let}
$A^{(k)}$, $\{ \Gamma_1,...,\Gamma_n \}$ {\it
be as in Theorem 1. Then the set}
$$
\{ A^{(k)}= \sum_{j} \alpha_j {\cal D}^{-1}|_{(k)}(x-x_j) ~|~ \{ x_j 
\}
\cap (\Gamma_1 \cup ... \cup \Gamma_n) \neq \emptyset  \}
$$
{\it is of measure zero.}\\

{\bf Lemma 2}\\
{\it
Let $\{ \Gamma_1,...,\Gamma_n \}$; $\eta$; $(\tau, \tau ')$
be as in Proposition 1;
${\cal D} \in
Cov(\tau , \tau ')$ be such that as in (\ref{spec}). Then}
$$
\hbox{Pr}\{ ~A^{(k)} = \sum_j \alpha _j {\cal D}^{-1} (x-x_j)~~ |~~
\overline{\lim _{n \rightarrow \infty} } \sum_{n \leq |x_j | \leq n+1}
|\alpha_j||{\cal D}^{-1}|_{(k)}(x-x_j)|> 0  \} =0
$$

In particular it follows that for any fixed configuration $\{
\Gamma_1,...,\Gamma_n \} \subset C_k$ the function (called $k$-
cocycles
Schwinger function)
$$
S(\Gamma_1,...,\Gamma_n) \equiv {\bf E} \hbox{e}^{i 
A^{(k)}(\Gamma_1)}...
\hbox{e}^{i A^{(k)}(\Gamma_n)}
$$
is well defined. For $k=1$, the corresponding $1$-cocycles Schwinger 
function
are known as Wilson loops (Schwinger) functions and as is well known 
they
play an important role in different physical theories, see i.e. 
\cite{Sei,
Bea} .
However, the almost sure results presented here are not very 
satisfactory 
due
to the problem of exceptional sets. To provide computable approach to
$k$-cocycles Schwinger functions $L^p$-version of the cosurface map
\ref{eq3.1}
should be given. We illustrate this in the case of random loop 
variables.

Let $\eta \in {\cal C}^{\infty}_0({\bf R}^4)$ be such that: $\eta 
\geq 0$,
supp$\eta \subset [-1,1]^{\times 4}$ and $\int \eta (x) dx =1$ and let
$\eta ^{\epsilon}(x) \equiv \epsilon ^{-4} \eta (\epsilon ^{-1}x)$.
For a given
loop $\Gamma \in C_1$ we define a family of test functions $\rho 
_{\Gamma
,k}^{\epsilon}(x) = \oint _{\Gamma} \eta ^{\epsilon}(x-z) dz^k$. Then 
we
define (the regularized) random loop variable:
$$
{}^{\epsilon}{\cal L}(\Gamma)=
\hbox{e}^{i (A_{(1)}, \rho ^{\epsilon}_{\Gamma})}
$$

{\bf Theorem 3}\\
{\it Let: $k=1$; $\eta$ be a regular $\tau$-covariant Poisson noise 
as 
above.
Let $\tau '$ be such that $\tau ' \supset \tau_{(1)} $  and let a 
regular
${\cal D} \in Cov(\tau , \tau ')$ be given. Assume that: }\\
(1) $|{\cal D}^{-1}_{kl}(x)| \leq \frac{c_{kl}}{|x|^{3+\epsilon 
_{kl}}}$
{\it for} $c_{kl >0}$, $\epsilon _{kl} >0$\\
(2) $|\int d \lambda (\alpha) (\hbox{e}^{<\alpha, y>}-1)|
\leq c |y|^{1+{\eta}}$ {\it for} $|y| \rightarrow 0$, $\eta \in
(\frac{1-min(\epsilon_{kl})}{3+min(\epsilon_{kl})},1]$.\\
(2') $|\int d \lambda (\alpha) (\hbox{e}^{<\alpha, y>}-1)|
\leq c |y|^{1+{\eta}}$ {\it for} $|y| \rightarrow +\infty$, $\eta \in
(-1, \frac{-max(\epsilon_{kl})}{3+max(\epsilon_{kl})})$.\\
{\it Then for any collection $\{\Gamma_1,...,\Gamma_n \}$ of loops 
and $p 
\in
[1, + \infty)$ there exists (in the norm $L^p(d \mu _{A})$sense) limit
$\lim_{\epsilon \rightarrow 0^+}
\prod _{l=1}^n {}^{\epsilon}{\cal L}(\Gamma_l)$ and then: }
$$
S(\Gamma_1,...,\Gamma_n) \equiv \lim_{\epsilon \rightarrow 0^+} {\bf 
E}
\prod _{l=1}^n {}^{\epsilon}{\cal L}(\Gamma_l)=
$$
$$
=\hbox{exp}\int dx \int d\lambda(\alpha) \{ \hbox{exp} (i \sum_{l=1}^n
\oint_{\Gamma_l} <\alpha , {\cal D}^{-1}(x- \cdot)>)-1 \}
$$
{Morever, the functionals $S(\Gamma_1,...,\Gamma_n)$ obey
a system of axioms as proposed in \cite{Sei}
with the exception of reflection
positivity.}

{\bf Remark}\\
{\it In the context of reflection positive Wilson loops functions 
suitable
technique for the reconstruction of quantum mechanical dynamics in 
the real
(Minkowski) time out of them was presented in \cite{Sei}.
The interesting problems
here are to find a convienient substitutes of Laplace-Fourier 
property and
HSSC in the context of not-necessarily reflection positive $k$-
cocycles
Schwinger functions that enables us to describe the corresponding 
real time
dynamics. }

{\bf Acknowledgements}\\
Essential part of this report has been prepared by one of the authors 
(R.G.)
during his stay at BiBoS Research Center, Bielefeld.


\end{document}